**Precipitation extremes and depth-duration-frequency under internal climate variability**
*Reducing the irreducible hydroclimate uncertainty*

**Udit Bhatia[1,2], Auroop Ratan Ganguly[1‡]**

[1]Civil and Environmental Engineering, Northeastern University, Boston, Massachusetts, United States, 02115
[2]Civil Engineering, Indian Institute of Technology, Gandhinagar, Gujarat, India, 382355

[‡]Corresponding Author Contact: Sustainability and Data Sciences Lab, Department of Civil & Environmental Engineering, 400 Snell Engineering, Northeastern University, 360 Huntington Avenue, Boston, MA 02115, USA; Phone: +1-617-373-6005; Email: a.ganguly@neu.edu

**Precipitation Extremes, Internal Variability, Extreme Value Theory, Depth-Duration-Frequency**

## ABSTRACT

Natural climate variability, captured through multiple initial condition ensembles, may be comparable to the variability caused by knowledge gaps in future emissions trajectories and in the physical science basis, especially at adaptation-relevant scales and projection horizons. The relations to chaos theory, including sensitivity to initial conditions, have caused the resulting variability in projections to be viewed as the irreducible uncertainty component of climate. The multiplier effect of ensembles from emissions-trajectories, multiple-models and initial-conditions contribute to the challenge. We show that ignoring this variability results in underestimation of precipitation extremes return periods leading to maladaptation. However, we show that concatenating initial-condition ensembles results in reduction of hydroclimate uncertainty. We show how this reduced uncertainty in precipitation extremes percolates to adaptation-relevant-Depth-Duration Frequency curves. Hence, generation of additional initial condition ensembles therefore no longer needs to be viewed as an uncertainty explosion problem but as a solution that can lead to uncertainty reduction in assessment of extremes.

# Introduction

Understanding and projecting precipitation, especially at regional scales, has traditionally been considered one of the "real holes" in climate science[1]. On the other hand, changes in precipitation extremes are acknowledged to be one of the most important climate change consequences relevant for impacts and adaptation[2]. Meanwhile, the lack of consensus about historical and future changes in precipitation mean and extremes, or our ability to extract signals of human activity, appears to be growing in the recent literature. One recent study [3] based on a 1940-2009 global land-based database of annual precipitation found little or no change in around 76% of the global land surface, while another [4] found support for the physical arguments that the seasonality of US daily heavy precipitation trends from 1979-2013 are linked more to internal decadal ocean variability than anthropogenic climate change. Conversely, one article [5] found that compared to 1950-1999, anthropogenic changes in precipitation means are already extractable in 2000-2009 in 36-41% of the globe, increasing to 68-75% in 2050 and 86-88% in 2100, while another[6] detected and attributed the influence of global warming on heavy precipitation in 41% of the observed areas across the globe, and a third article [7] found statistically significant increasing trends precipitation extremes in about two-thirds of stations across the globe and a statistically significant association with globally averaged near surface temperature at a rate between 5.9% and 7.7% per Kelvin. While it is certainly possible that each study is accurate based on the specific choice of precipitation extremes metric(s), dataset(s) and analysis method(s) of choice, the overall picture remains unclear.

This lack of clarity is exacerbated by internal climate variability (ICV), which is in turn manifested through multiple initial condition ensembles (MICE). Projections from global climate or earth system models (ESM) suffer from three broad class of uncertainties: (a) knowledge gaps in how coupled human and natural systems may lead to greenhouse-gas emissions scenarios as captured through representative concentration pathways (RCP) which drive the ESMs, (b) lack of understanding of the physics and biogeochemistry of the climate and earth system the variability of which is encapsulated through structural and parametric differences in multiple model ensembles (MME), and (c) the intrinsic variability of the climate system which is captured as extreme sensitivity to initial conditions through MICE, which in turn are developed for a specific RCP and a specific model within the MME. While RCP variability is based on emissions and hence controllable by humans and treated as what-if scenarios for policy, MME have been handled statistically ranging through methods ranging from simple model weighing strategies to skill-based selection that may attempt to balance historical skills when compared with observations with future consensus among models as well as with the goodness of the embedded science understanding. However, ICV correspond to what has been called [8,9] Irreducible uncertainty, which is expected to be large at local and even regional scales. However, while ICV may further obscure any signals of change in mean or extremes of precipitation or in our ability to attribute such change to human drivers. Thus, three studies[9–11] have shown that local and regional precipitation trends are likely to remain within the bounds of internal climate variability over most of the globe throughout the 21st-century even though spatially aggregate projections of precipitation extremes may be relatively more robust[9]. However, a large part of the local to regional differences in trends of extremes were found to be explained by ICV, which was found to regionally obscure or amplify the forced long-term trends for many decades[11].

While researchers have examined model-simulated precipitation extremes with a physical diagnostic that distinguishes between thermodynamic and dynamic drivers[12–16] and suggested that recent precipitation extremes observations confirm theory and models[17], clear lack of consensus exists about the signal of change or attribution that can be extracted despite the natural or intrinsic variability [5,18] as well as how the extracted signals may be explained[19]. The ability of ICV to mask trends in precipitation, both across the globe[20] and in the US [21], as well as the need to explain this irreducible uncertainty to stakeholders[22] have been explained in the literature. The case has also been made[21,21,22] for why more MICE runs may be needed beyond the current generation of large ensemble runs and why a balance may be needed between increasing the number of MICE runs needed to fully explore the space of plausible futures versus enhancing the resolution of models. The need for large numbers of MICE runs for a comprehensive characterization of the irreducible uncertainty, as well as the questions around data (ensemble) sufficiency, may lead to a situation where comprehensive assessments of uncertainty may be even more difficult owing to the multiplier effects of ensembles.

**Handling uncertainty from Multiple Initial Condition Ensembles**

The nature of the irreducible climate uncertainty and the potential explosion of model runs (i.e., MICE generated from multiple models with multiple emissions scenarios) leads to a set of research questions that animates our work here. Specifically, here we examine, first, how fundamental is a comprehensive characterization of ICV to our understanding of hydroclimate science and how relevant is it for informing stakeholders interested in adaptation or policy Second, we need to understand if a strategy demonstrated in Thomson et al 2017 and Thomson et al 2019 [23,24], in the context current climate scenarios can be used to make sense of the larger number of GCM model ensembles that will become available, especially MICE runs in the context of ICV. Hence, The premise of our analysis rests on a simple hypothesis used in[23,24], which can simultaneously mitigate the ensemble multiplier effect for ICV while constraining uncertainty in return period curves, in this case for precipitation extremes. In the context of precipitation extremes under internal variability, multiple initial condition ensembles (MICE) from a given model essentially encapsulate the same physics including identical structural and parametric forms, the simulations provide different plausible versions of the real world.

As a consequence, the statistics of the runs may be estimated by considering multiple (MICE) runs at once. Thus, if *N* MICE ensembles are available, each with *M* projection points (e.g., space-time buckets, representing daily values for *Y* years, say), the all the *N*×*M* points may be used to estimate statistics like the overall mean and sample standard deviation, while the lag-1 sample autocorrelation may be estimated from all *(M–1)* pairs of adjacent points in each of the *N* ensembles, or a total of *N*×*(M–1)* samples. Thus, instead of having *N* estimates of mean, standard deviation and lag-1 autocorrelation or other statistics, each with estimation uncertainties, leading to the possibility of explosion of uncertainties across (MICE) ensembles, the *N* time series can in a sense be concatenated to generate single estimates of the statistics, where the larger sample size of the estimates can lead to reduced uncertainties. While temporal correlation length scales may need to be considered in certain estimation problems, in situations where independent and identical distributions (IID) are assumed, such considerations are not necessary.

These considerations can be rather useful for estimates based on extreme value theory (EVT) such as the (*T-year*) return levels of precipitation extremes, corresponding to a probability of occurrence of $T^{-1}$ in any given year, since data are limited for extremes and estimation uncertainties can be reduced with additional data. Thus, one way to estimate the *T-year return level* (RL$_T$) of precipitation extremes (methods in [14,25,26]) from time series of precipitation is to extract high values (specifically, annual maxima) which may be assumed to be IID (as assumption that needs to be tested for statistical significance), then fitting EVT distributions (e.g., Generalized Extreme Values [GEV]) to these values, and finally inferring extreme value properties (e.g., 100-year return levels, or RL$_{100}$) from the parameters of the fitted distributions [14,27]. If our hypothesis about identical data generation process across MICE ensembles is true and hence the consequences in the estimation processes as discussed above are valid, then uncertainty in the estimates of RL$_{100}$ can be reduced by concatenation of the MICE time series.

One aspect of the above is that instead of having to view MICE as contributors to the explosion of uncertainties owing to the ensemble multiplier effect (RCP ensembles times MME times MICE), they can be viewed as generators of additional samples which will likely reduce the uncertainties in our estimates. This is especially valuable for extremes analysis since sample size is relatively even more critical to estimation uncertainties given the rarity of the extremes[27]. Furthermore, unlike in certain situations where the plausible futures projected by each ensemble member may be useful to consider for adaptation, in the case of precipitation extremes it is statistics such as $RL_{100}$ and depth-duration-frequency (DDF) curves which would influence adaptation choices in hydraulic engineering design or operations and water resources planning or management.

## Methods and Datasets

**Datasets**

We use outputs from NCAR CESM Large Ensemble project (LENS) [20,28], which are intended to advance understanding of internal variability in changing climate. LENS was generated with the NCAR Community Earth System Model run at ~1° horizontal resolution. The ensemble methodology branches multiple GCM simulations from a single CMIP-type transient Historical Simulation. These multiple realizations differ only in slight perturbations in the initial atmospheric conditions in 1920. Each ensemble member is then prescribed the transient historical forcing through the end of the CMIP5 historical period (2005), and the Representative Concentration Pathway (RCP 8.5) transient forcing after 2005. For this study, we use large scale precipitation, convective precipitation and average annual surface temperature from each MICE to test and explore mechanistic plausibility of the hypothesis. LENS datasets can be downloaded from: http://www.cesm.ucar.edu/projects/community-projects/LENS/

To compare the estimation of extremes as obtained from LENS with observations, we use Climate Prediction Center (CPC) unified gauge-based analysis of Daily Precipitation at the resolution of 0.25 degrees. Observed precipitation is first interpolated to match the resolution of CESM LENS outputs. Since, simulations from climate modes require bias correction prior to use in the impact assessments, we bias correct the historical runs (from the period of (1948-2005) of Multiple Initial Condition Ensembles (MICE) using quantile mapping [29] which is routinely used to biases of global and regional climate models, compared to observational data. Observation data can be freely downloaded from : https://www.esrl.noaa.gov/psd/data/gridded/data.unified.daily.conus.html

To compare the contribution of multiple model and multiple initial condition runs to variability in extremes, we use the outputs from 27 Model Ensembles (MMEs) from Climate Model Intercomparison Project (CMIP5) listed in table S1 and same number of initial conditions (MICE) [IC2-IC28] under 8.5 RCP transient forcing. Since, we are comparing the statistics of multiple model ensembles, we use same number of samples from each ensemble to avoid sampling bias. MICE data listed in table S1 is freely available from:
https://cmip.llnl.gov/cmip5/data_portal.html

**Methodology**

Based on prior literature[14,25,27,30], we use Generalized Extreme Value theory (GEV) distribution for estimation of return levels of precipitation events. GEV, which is based on the block maxima theory is utilized to quantify the intensity of extreme precipitation. Given the duration of interest (e.g. 1-day data), the Annual Maximum Precipitation (AMP) series are computed from data; these are then used to estimate parameters of fitted GEV. In many studies, extreme rainfall statistics are expressed in terms of $T$ year rainfall depth that statistically represents one annual maximum precipitation event exceeding a given threshold is expected to occur within every non-overlapping T year length window. Details of GEV can be found in [27] and detailed methodology in the context of climate models is discussed in [14,26,30]. The GEV parameters are estimated using the maximum likelihood estimates. We use *fevd* and *eva* library in R to estimate return levels, 95% confidence interval around the estimate and generate random samples from extreme value distribution with estimated parameters for testing statistical significance. We test for the goodness of fit of the extreme value distributions using Kolmogorov-Smirnov at a 5% significance level. If a grid-point fails to pass the goodness of fit test, it is not included in the computation of regional statistics. Less than 1% of the total cases fail to pass the goodness of fit tests suggesting the appropriateness of the GEV distribution in modeling precipitation extremes. In concatenation approach, we use annual maximum precipitation from each year and concatenate these together. For example, 27 initial conditions with 96 years of daily data will yield (27x96) AMPs, which in turn, are used to estimate return levels and associated 95% confidence interval bound around the estimates.

To understand the significance of difference between median and mean values in MME and MICE for each of the 9 regions and entire US (Figure S1), we use Kruskal Wallis H-test (non-parametric) and 2-sided independent T test (parametric) at significance level of 0.05 to test whether average value of variability and return levels estimated from MME and MICE differ significantly in each region. We observe p-values <<0.05 for IQR as well as values of 100-year return level for both mean and median values. Thus, we can reject the null hypothesis of equal averages form IQR and RL100. We use the period of 2005-2100 from RCP8.5 scenario. for both MME and MICE.

To illustrate the continuous change in statistics of extreme precipitation and contribution of natural variability in overall variability, we account for non-stationarity by using moving window of 30-years to estimate return levels from Multiple Initial Condition Ensembles as well as Multiple Model Ensembles (Figure S2). To address the potential non-stationary effect, rather than using all temporal data to fit one GEV distribution, GEV parameters are estimated separately for 30-year moving windows. The 30-year size is selected as it is expected to smooth out the effects of most climate oscillators and also give more confidence for low frequency extremes. We notice that even with non-stationarity, MME variability dominates than that obtained from MICE across all time-horizons. However, values of 100-year return levels (RL-100) obtained from MME are significantly higher for entire CONUS as well as 9 regions under consideration.

Rather than focusing exclusively on daily precipitation, we also compute return levels for cumulative rainfall for the period of 1,2, 3,5,6, and 10 days. Thus, AMP series are computed to derive return level corresponding to 100-year and 30-year precipitation levels. These extreme rainfall estimates and corresponding 95% CI are illustrated in terms of Depth Duration Frequency (DDF) curves, which are commonly used in for design and management of hydraulic structures and critical infrastructure facilities.

We explore the mechanistic justification behind the ability to concatenate MICE by measuring variability in the fraction of precipitation extremes generated from convective (versus large scale) processes across latitudinal bands. We use interdependent T-test to measure the statistical significance of average uncertainty bound at significance level of 0.05. Since, p>0.05, null hypothesis that mean values are significantly different cannot be rejected in the case of MICE. However, for MME, p<<0.05 suggests that upper and lower bounds are significantly different. All significance tests are performed using *scipy* package in Python.

To explore the scaling relationships between precipitation extremes and annual average surface temperature for MICE, we measure the changes in return levels ($RL_{100}/T$) and heavy (99.9-percentile: $P_{99.9}$) precipitation ($P_{99.9}/T$). The 99.9th percentile of daily precipitation (mm/day) is computed for the periods of 2081-2100 and 1981-2000 is used a reference period to calculate interquartile range at each latitude band. We chose these specific time-periods to heavy precipitation events so that scaling relationships observed in MICE can be compared to those of MMEs reported by Gorman et al.[15].

# Results

Figure 1 contrasts the variability in $RL_{100}$ over the 21st-century within the Continental United States (CONUS) between MME (27-member model ensembles from CMIP5; *See* SI) versus MICE (27-member ensembles from NCAR LENS). While the variability across MICE is statistically significantly less than that across MME, ignoring MICE variability can result in significant underestimation of $RL_{100}$ for the US overall and across the nine hydrometeorological zones. The adaptation-relevance is shown by plotting the Depth-Duration-Frequency (DDF) which informs water resources and hydraulic infrastructures management. While the implication of ignoring DDF are evident across the CONUS, in certain zones such as the US Northeast $RL_{100}$ levels may be underestimate to $RL_{30}$ unless MICE are considered. We note that MICE variability here is based on one model only, thus the underestimation may be even more significant once and if MICE become available across models.

Figure 2 shows the validity and implications of our hypothesis described previously regarding concatenation of MICE ensembles. The aleatoric uncertainty in $RL_{100}$ computed from the average of the 27-member MICE reduces significantly following concatenation, which also is more narrowly constrained around the observed variability. The $RL_{100}$ based on the concatenated MICE are significantly closer to the observed $RL_{100}$ in all nine zones and over the CONUS with significantly narrower uncertainty bound. Since the outputs of GCMs suffer from systematic bias and scale dependent predictability[31], we apply a quantile mapping based bias correction to daily precipitation values simulated by the Global Circulation Models. Bias corrected values from various initial condition runs is then used to compare the return $RL_{100}$ estimated from observations with that of model outputs. We note that while Thompson et. al 2017[23] assessed the model fidelity using the agreement between mean, standard deviation, skewness and kurtosis computed from model output and observations, we use the return levels, which in turn is function of scale, location and shape factor to assess the agreement between observations and bias corrected outputs.

Figure 3 depicts the adaptation-relevance of the results in Fig. 1 based on DDF curves and their uncertainty bounds across the US hydrometeorological zones. Specifically, while the computed $RL_{100}$ and $RL_{30}$ and the corresponding DDF curves are not statistically distinguishable from each other when MICE members are considered individually, the concatenated MICE make the uncertainty bounds significantly narrower around each return level estimate and the DDF curves statistically significantly distinguishable from each other. It is noted that similar improvement in risk estimates of the chance of unprecedented rainfall was reported in. In present case, we report the reduced uncertainty for US hydrometereological zones in terms of engineering design and infrastructure management relevant intensity duration and frequency curves.

While Figure 1 establishes the importance of this study, Figures 2-3 establish the validity and the significance of the hypothesis as well as the benefits in terms of generating improved and potentially more actionable adaptation-relevant information. Figure 4 explores the mechanistic justification behind the insights in Figures 2-3. The rationale for our hypothesis about the ability to concatenate MICE was based on the notion that – unlike MME – the fundamental physics (including parametric and structural selections) and hence the data-generation process remains identical across MICE. Figure 4 shows that the variability in the fraction of the precipitation extremes generated from convective (versus large scale) processes across latitudinal bands are nearly identical (statistically indistinguishable from zero: See Methods) across MICE ensembles but significant across MME ensembles. This shows that the directly controllable physics and parameters in the MICE runs are nearly identical unlike in the MME runs. When the temperature dependence of changes in return levels ($\Delta RL_{100}/\Delta T$) or heavy (99-percentile: $P_{99.9}$) precipitation ($\Delta P_{99.9}/\Delta T$) are computed, the variability in MICE ensembles are not dissimilar to the corresponding MME-variability reported in the literature. This suggests that the physics that is an indirect result of the parameter choices (rather than being directly controllable) may still exhibit variability, possibly owing to how the differences in MICE initializations may end up manifesting through the modeling system (e.g., as an analogy, the impact of initialization on a simulated low-frequency climate oscillator is shown in the online supplement of [32]).

# Conclusion and Future Work

We have used a hypothesis-driven approach which (in essence) reduces what has been considered the irreducible uncertainty component in depth duration and frequency curves under internal climate variability. The approach can help manage the explosion of uncertainty from (one of contributors to) the ensemble multiplier effect. This is accomplished by exploiting each MICE-run as an additional sample-generator for the estimation of precipitation extremes statistics over a projection time-window rather than producing an estimate from each run which then needs to be reconciled. The proposed approach does not reduce the intrinsic climate variability (which may be indeed irreducible given the extreme sensitivity to initial conditions [33,34]), but instead of looking at MICE merely as manifestations of irreducible uncertainty, finds a way to use MICE runs to reduce the uncertainty in the estimates, in this case for precipitation extreme under a changing climate. The uncertainty reduction happens for two mutually related and reinforcing reasons: first, instead of producing as many estimates as the number of initial condition ensembles, we produce just one estimate, and second, the one estimate is produced with a much larger sample than each of the individual estimates. The ability of our proposed approach to improve our understanding of precipitation extremes under climate change and correspondingly generate more credible adaptation-relevant metrics is demonstrated as has been done for current climate elsewhere [23]. Furthermore, we provide a mechanistic justification of our hypothesis which forms the basis of this study while also pointing to the possible caveat. The irreducible uncertainty component and MICE simulations are likely to be considered more seriously by climate scientists and stakeholders in the future including in the upcoming IPCC AR6 and CMIP6, thus enhancing the implications of this work. The lessons learned from this work may be developed further for broader applications to extremes generated from nonlinear spatiotemporal dynamical systems across physical or biological sciences, engineering and social sciences. The lessons learned for precipitation extremes projections and the basic approach here may generalize to multiple climate variables and extremes as well as to more broadly to climate science understanding and adaptation-relevance.

# Acknowledgements

This work was supported in part by the Civil and Environmental Engineering Department, Sustainability and Data Sciences Laboratory, Northeastern University and Civil Engineering Department, Indian Institute of Technology, Gandhinagar. The work of U. Bhatia and A. R. Ganguly were supported by four National Science Foundation Projects, including NSF BIG DATA under Grant 1447587, NSF Expedition in Computing under Grant 1029711, NSF CyberSEES under Grant 1442728, and NSF CRISP type II under Grant 1735505. We are thankful to Evan Kodra, CEO, risQ Incorporated, Cambridge, MA, Kate Duffy, Northeastern University and Divya Upadhyay, Indian Institute of Technology, Gandhinagar for their helpful comments and suggestions.

## Author Contributions Statement

A.R.G., and U.B. designed the experiments. U.B. performed the experiments and analyzed the data. U.B. and A.R.G. wrote the manuscript.

## Competing Interests

The authors declare no competing interests.

## Corresponding author

Correspondence to A.R.G.

## Data Availability

The data that support the findings of this study are openly available at http://www.cesm.ucar.edu/projects/community-projects/LENS/, https://www.esrl.noaa.gov/psd/data/gridded/data.unified.daily.conus.html and https://cmip.llnl.gov/cmip5/data_portal.html

**Figure 1. Precipitation extremes may be underestimated, leading to misinformed adaptation, when internal climate variability is not considered.** **(A) (i-iii)** Analysis within Contiguous US showing $RL_{100}$ for three NCAR LENS CESM1 CAM5.2 runs. **(iv)** 27-model Inter Quartile Range (IQR) computed from MICE. **(v-vii)** same as **(i-iii)** but for three models from MME selected randomly. **(viii)** same as **(iv)** but for MME. **(B)** IQR from MME shows significantly larger variability (p-value<0.05) but significantly higher upper bounds are obtained from MME. **(C)** shows adaptation implications via DDF curves (e.g., in NE zone, $RL_{100}$ from MME compares to $RL_{30}$ from MICE). Three of nine US hydrometeorological zones are shown, the rest are in SI (Fig. S2-S3). The results demonstrate the criticality of ICV to informing the design or operations of hydraulic infrastructures and water resources planning or management.

**Figure 2. Concatenating MICE, justifiable owing to identical physics, can reduce the so-called irreducible uncertainty.** MICE (unlike MME) is assumed to be driven by identical physics, while extreme value theory (EVT) assumes IID of annual maxima, hence MICE can be concatenated to increase the sample size per estimate and reduce the number of estimates that need to be reconciled. **(A)** shows reduction of uncertainty in $RL_{100}$ estimates within CONUS with 95% CI (right: average of 27 MICE; left: concatenated); **(B)** shows kernel density estimates for the CONUS' **(C)** shows agreement between $RL_{100}$ estimated from concatenated MICE and observations across nine US hydrometeorological regions. The reduction of uncertainty is a result of larger samples and lesser number of estimates to reconcile in the final estimate of precipitation extremes $RL_{100}$, the parameter that informs adaptation, even though the uncertainty from the ICV (extreme sensitivity to initial conditions) per se cannot be reduced.

**Figure 3. Reducing the irreducible uncertainty for impact and adaptation relevant metrics.** Uncertainty in Depth-Duration-Frequency (DDF) curves, which are in turn often used to guide the design, maintenance and operations of hydraulic infrastructure such as dams, levees and reservoirs, and for water resources planning, are reduced through MICE concatenations as seen for three US hydrometeorological regions (other six regions in SI: Fig. S3): Large uncertainties in MICE leads to overlaps confidence bounds of $RL_{100}$ and $RL_{30}$ estimates **(A)** while reduction of uncertainties (see Fig. 2) makes **(B)** the $RL_{100}$ significantly distinguishable from $RL_{30}$. The reduction of irreducible RL-estimates leads to statistically significant delineation of adaptation-relevant DDF.

**Figure 4. Mechanistic rationale and statistical justification of uncertainty reduction hypothesis.** The parameters and structural forms are identical for MICE runs unlike in MME, which manifests as nearly identical IQR with $25^{th}$ and $75^{th}$ percentiles being statistically indistinguishable (p>0.05) for all latitudes for MICE **(A)** but not so for MME **(B)** when the percent of convective to total precipitation is examined over the CONUS. However, while the physics that are more directly controlled by the parameter choices may be identical, variability persists in the physics that is indirectly determined, such as temperature dependence of precipitation extremes, whether $RL_{100}$ **(C)** or 99.9 percentiles **(D)**. Uncertainty reduces continuously via successive concatenations, as shown in **(E)** with no concatenation to full concatenation of all available MICE runs (randomly-selected concatenated time series are shown as illustrations when concatenation is between 1 and 27), both owing to large sample-sizes per estimate as evident from **(E)** and lesser number of estimates to reconcile as evident from the box plots of the upper and lower bounds of the $RL_{100}$ estimates from each of the 27 MICE runs versus the estimates from the fully concatenated time series **(F)**.

**Figure 1**

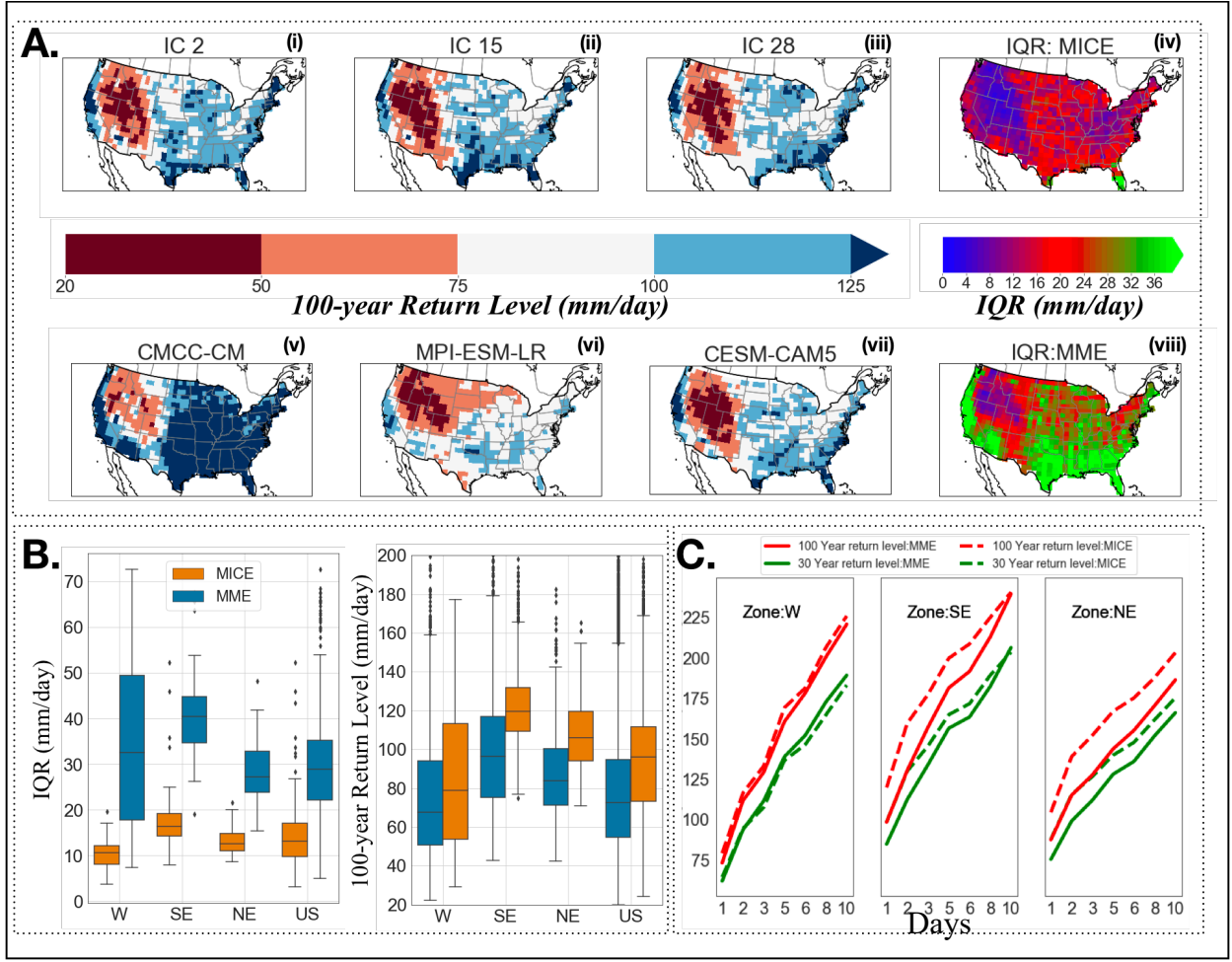

**Figure 2**

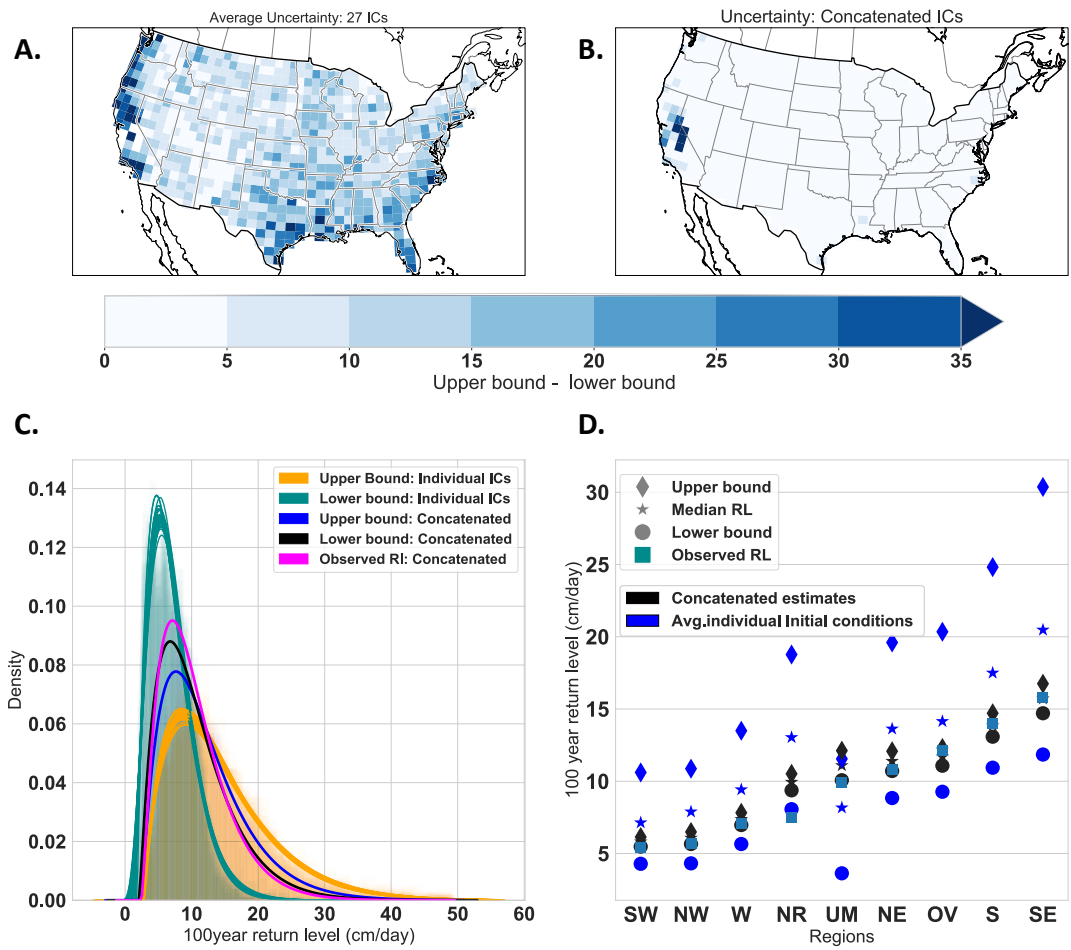

**Figure 3**

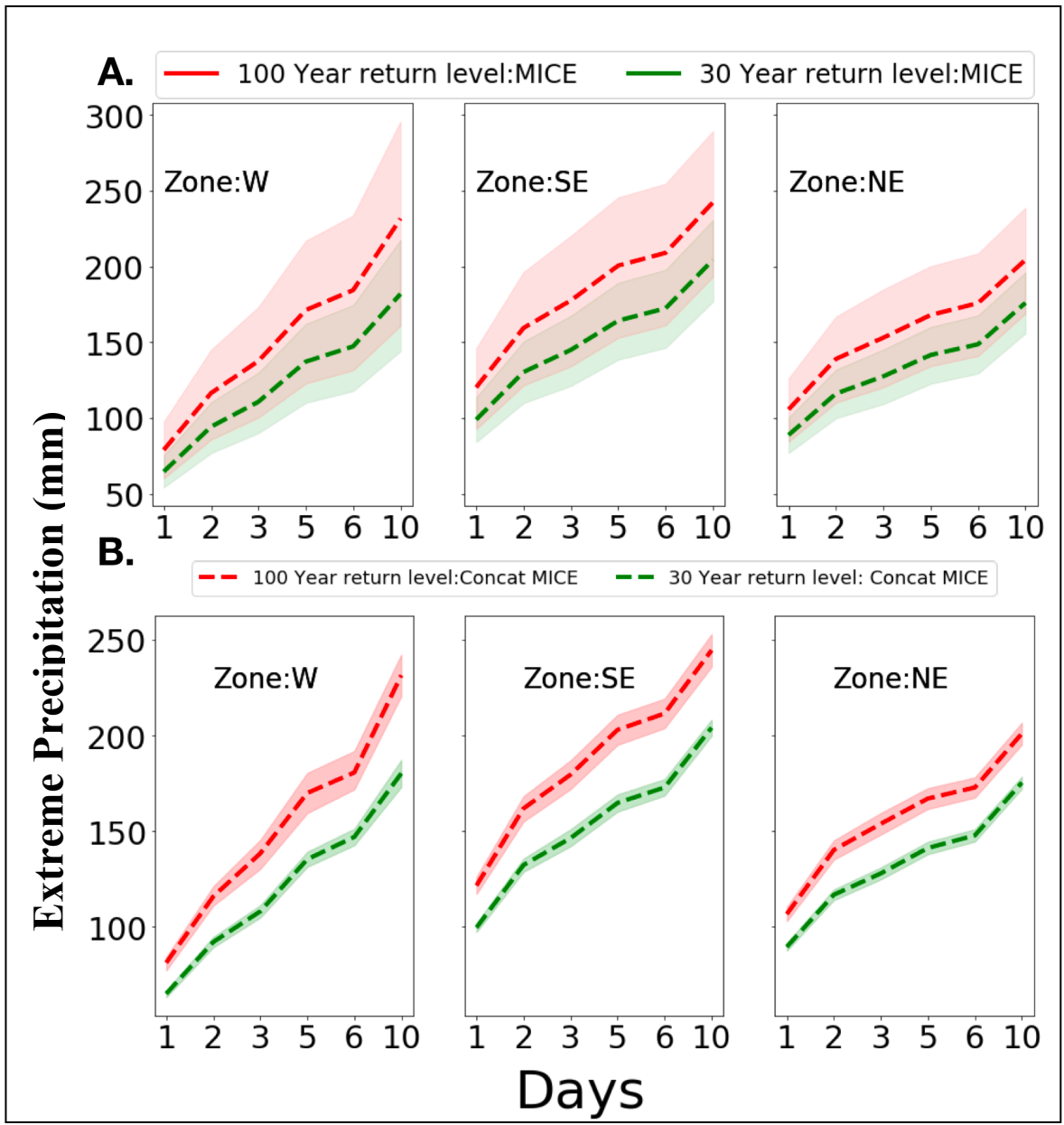

**Figure 4**

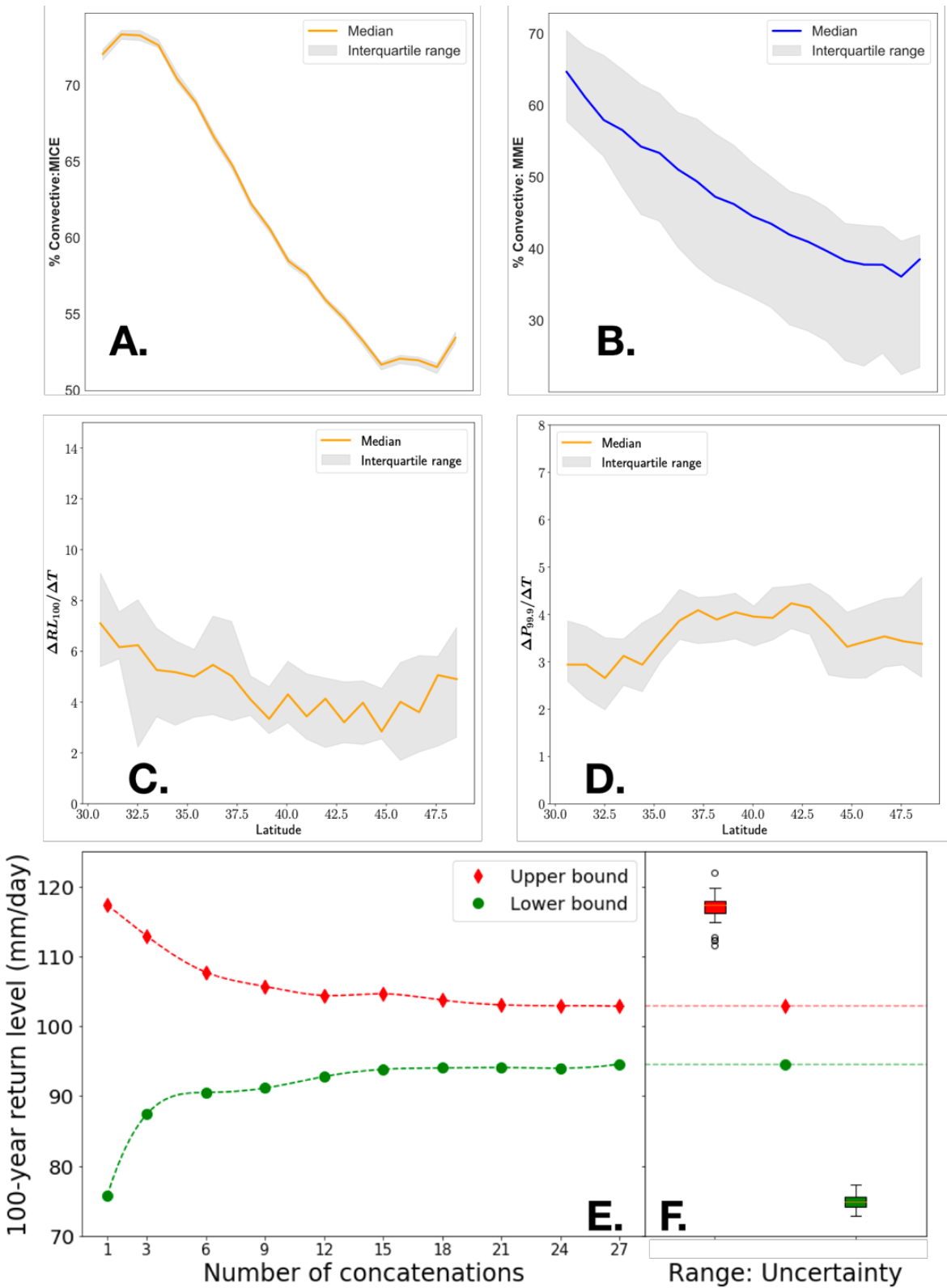

**Supplementary Information**
**Precipitation extremes and depth-duration-frequency under internal climate variability**
*Reducing the irreducible hydroclimate uncertainty*

Udit Bhatia, Auroop Ratan Ganguly

**SI: Precipitation extremes and depth-duration-frequency under internal climate variability**

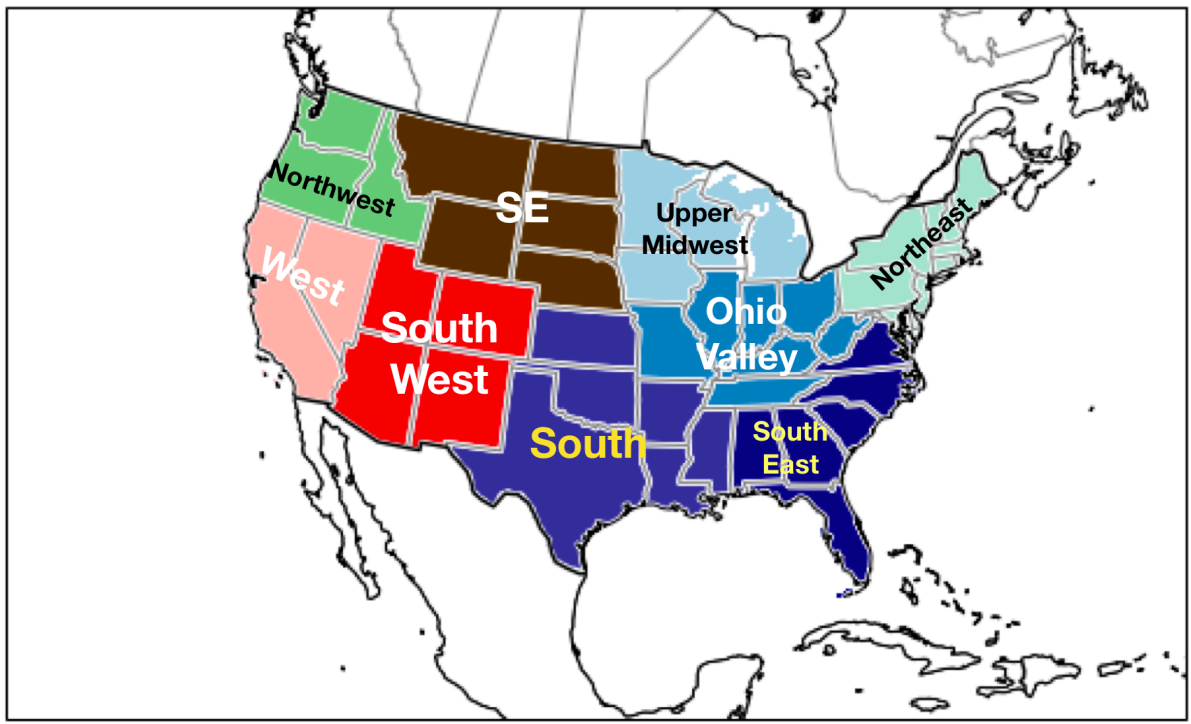

**Fig. S1:** Nine climatically consistent regions within the contiguous United States which are useful for putting current climate anomalies into a historical perspective have been identified by National Centers for Environmental Information scientists. Figure above shows these 9 regions. Figure generated using Python 3.6 and shapely[1]

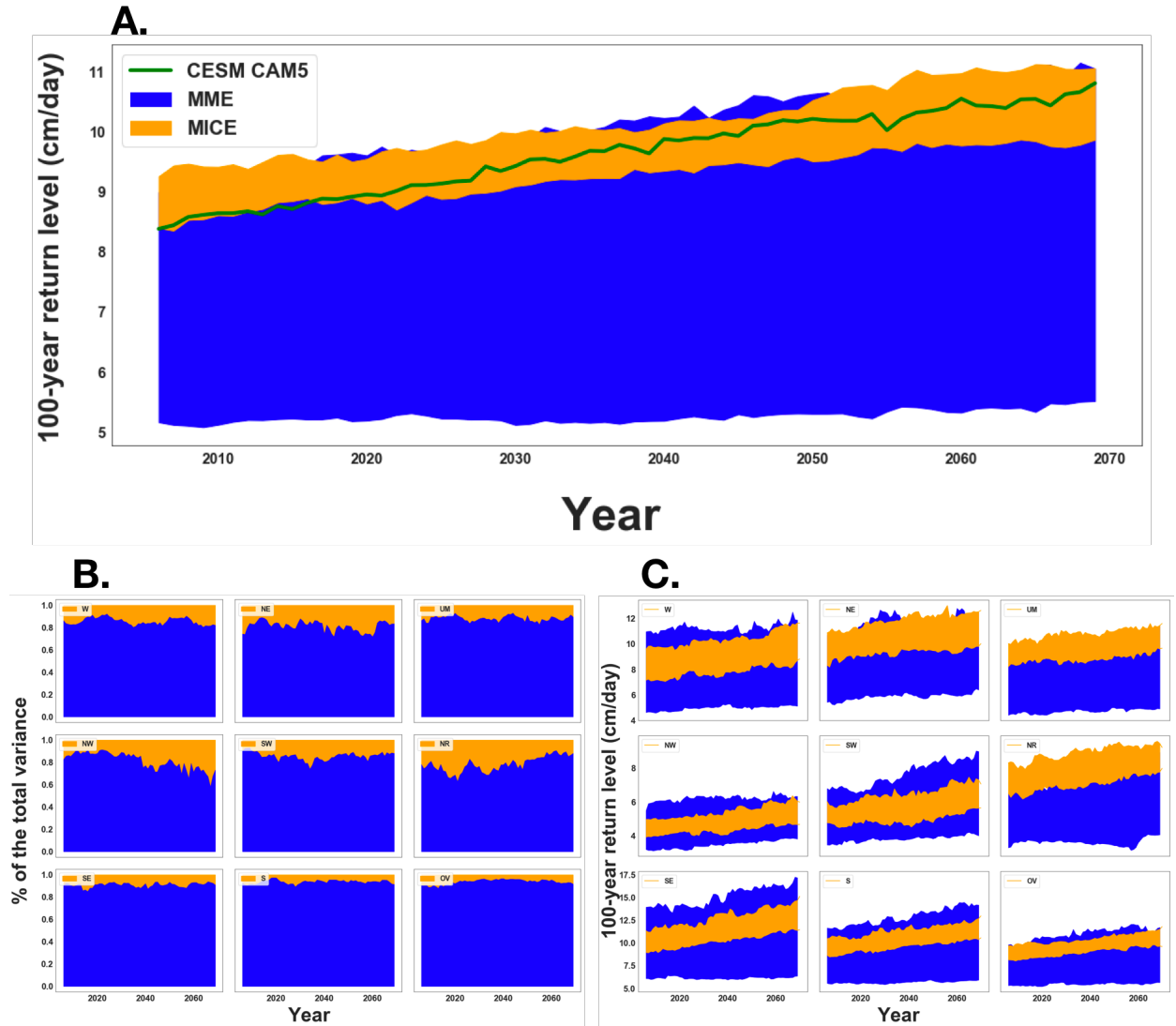

**Fig. S2: A.** Average estimates of 100-year return level (RLs) for entire CONUS using 30-year moving window beginning at 2006. Green line shows the RLs as obtained from CESM-CAM5, which is one of the ensembles in MME **B.** The ratio of MICE to MME variability is not significantly different across climate time horizons over this century for the 9 regions **C.** Same as figure A but for 9 climatological regions (See S1).

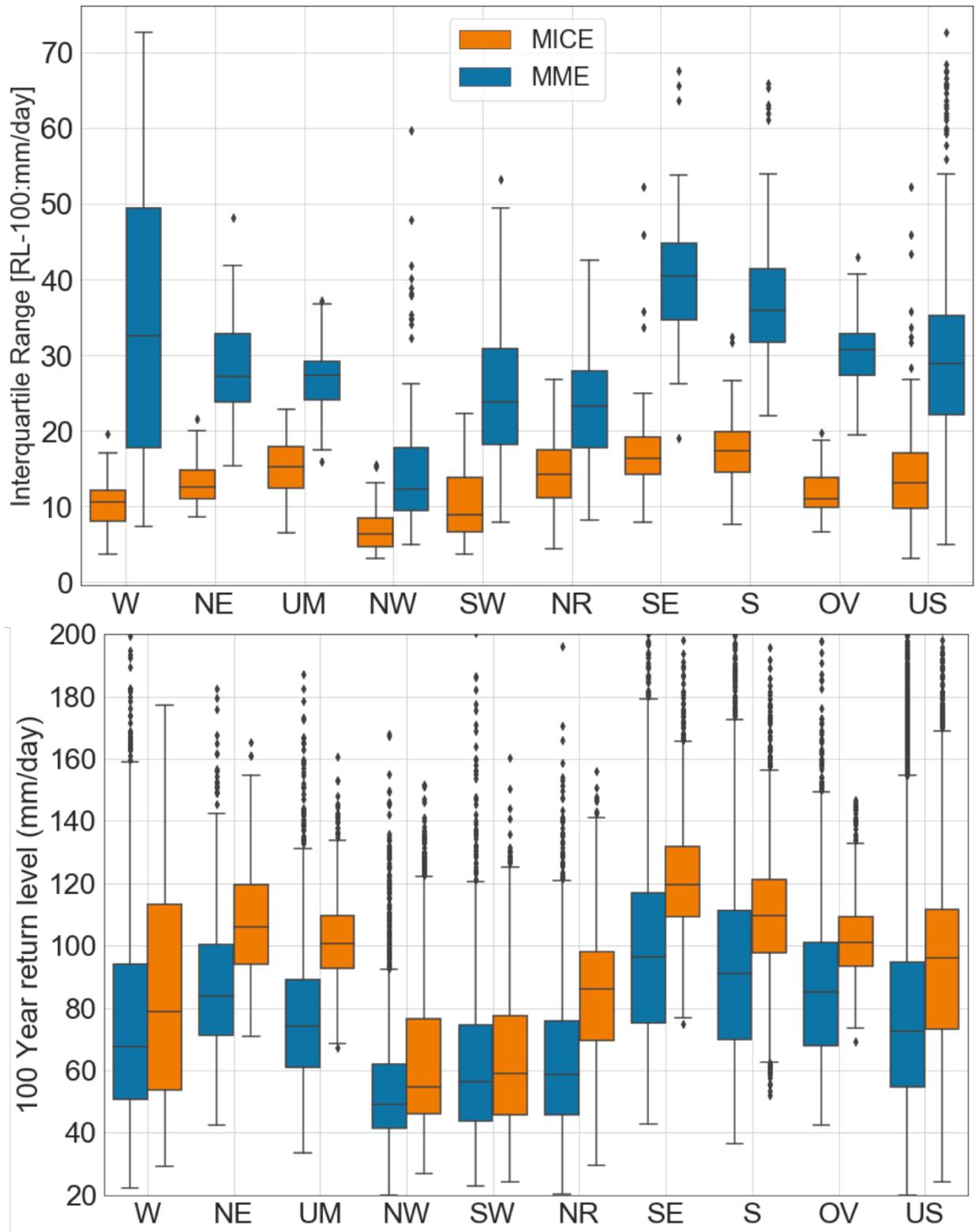

**Fig. S3**: Same as Figure 1B but for 9 regions shown in S1.

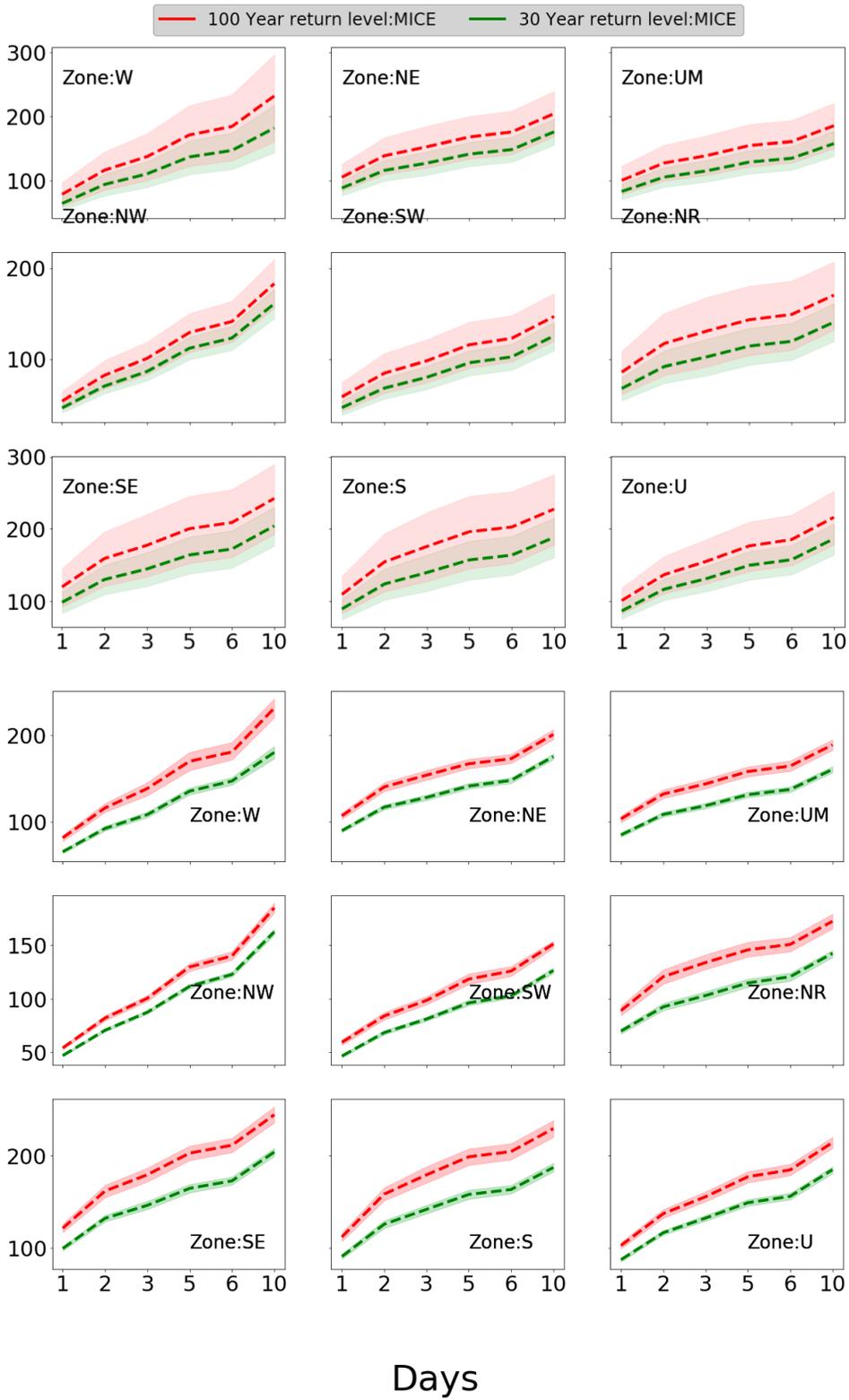

**Figure S3:** Same as Figure 3 but for all 9 regions shown in S1

**Table S1: List of CMIP5 models used in this study.**

| Modeling Group | Model Name |
|---|---|
| Commonwealth Scientific and Industrial Research Organization (CSIRO) and Bureau of Meteorology (BOM), Australia | ACCESS1-0 |
| | ACCESS1-3 |
| Beijing Climate Center, China Meteorological Administration | bcc-csm1-1 |
| | bcc-csm1-1-m |
| Beijing Normal University | BNU-ESM |
| Canadian Centre for Climate Modelling and Analysis | CanESM2 |
| National Center for Atmospheric Research | CCSM4 |
| | CESM1-BGC |
| | CESM1-CAM5 |
| Centro Euro-Mediterraneo per I Cambiamenti Climatici | CMCC-CESM |
| | CMCC-CM |
| | CMCC-CMS |
| Commonwealth Scientific and Industrial Research Organization in collaboration with Queensland Climate Change Centre of Excellence | CSIRO-Mk3-6-0 |
| EC-EARTH consortium | EC-EARTH |
| State Key Laboratory Numerical Modeling for Atmospheric Sciences and Geophysical Fluid Dynamics | FGOALS-g2 |
| NOAA Geophysical Fluid Dynamics Laboratory | GFDL-CM3 |
| | GFDL-ESM2G |
| | GFDL-ESM2M |
| Institute for Numerical Mathematics | inmcm4 |
| Institut Pierre-Simon Laplace | IPSL-CM5A-LR |
| | IPSL-CM5A-MR |
| | IPSL-CM5B-LR |
| Atmosphere and Ocean Research Institute (The University of Tokyo), National Institute for Environmental Studies, and Japan Agency for Marine-Earth Science and Technology | MIROC-ESM |
| | MIROC-ESM-CHEM |
| | MIROC-5 |
| Max Planck Institute for Meteorology | MPI-ESM-LR |
| | MPI-ESM-MR |
| Meteorological Research Institute | MRI-CGCM3 |
| Norwegian Climate Centre | NorESM1-M |